\documentclass[twocolumn,prc,showpacs,amsmath]{revtex4}

\usepackage{epsfig}
\usepackage{graphicx}% Include figure files
\usepackage{dcolumn}% Align table columns on decimal point
\usepackage{bm}% bold math
\usepackage{graphics}
\begin{document}

%\preprint{APS/123-QED}

\title{ANALYTICAL FORM OF THE DEUTERON WAVE FUNCTION
CALCULATED WITHIN THE  DISPERSION APPROACH}

\author{A.F.~Krutov}

\affiliation{Samara State University, 443011 Samara, Russia}
\email{krutov@ssu.samara.ru}

\author{V.M.~Muzafarov}
\affiliation{Steklov Mathematical Institute of the Russian Academy of Sciences, Moscow
119991, Russia}
\email{victor@mi.ras.ru}

\author{V.E.~Troitsky}
\affiliation{D.V.~Skobeltsyn Institute of Nuclear Physics,
  Moscow State University, Moscow 119992, Russia}
\email{troitsky@theory.sinp.msu.ru}
\date{\today}

\begin{abstract}
We present a convenient analytical parametrization of the
deuteron wave function calculated within dispersion approach as a
discrete superposition of Yukawa-type functions, in both
configuration and momentum spaces.
\end{abstract}


\maketitle

Recently in the paper ~\cite{KrT06} it was shown that the deuteron
tensor polarization component ${T_{20}(Q^2)}$ provides a crucial
test of deuteron wave functions in the range  of momentum
transfers available in to-day experiments. The calculation
~\cite{KrT06} shows that the most popular model wave functions do
not give adequate description of $T_{20}(Q^2)$ and are to be
discriminate in favor of those obtained in the dispersion
potentialless inverse scattering approach with no adjustable
parameters \cite{MuT81}, (see \cite{Tro94}, too) and giving the
best description. Some time ago this function (MT - wave
function) was used in the calculation of the neutron charge form
factor~\cite{KrT03}. The results of calculation (12 new points)
are compatible with existing values of this form factor of other
authors. A fit is obtained for the whole set (36 points) taking
into account the data for the slope of the form factor at $Q^2 =
0$. These results will be used in the neutrino scattering
experiments in Fermilab \cite{DrS04}.

The aim of the present paper is to present a conventional algebraic
parametrization of the deuteron MT-wave function
calculated within dispersion approach as a discrete superposition of
Yukawa--type terms.

Let us remind briefly the main characteristic features of these wave
functions obtained in the frame of the potentialless approach to the
inverse scattering problem.

The important feature of these wave functions is the fact that
they are "almost model independent":  no form of $NN$ interaction
Hamiltonian is used. The MT wave functions are given by the
dispersion type integral directly in terms of the experimental
scattering phases and the mixing parameter for $NN$ scattering in
the $^3S_1-^3D_1$ channel. Regge--analysis of experimental data
on $NN$ scattering was used to describe the phase shifts at large
energy.

It is worth to notice that the MT wave functions were obtained
using quite general assumptions about analytical properties of
quantum amplitudes such as the validity of the Mandelstam
representation for the deuteron electrodisintegration amplitude.
These wave functions have no fitting parameters and can be
altered only with the amelioration of the $NN$ scattering phase
analysis.

Let us notice that the process of constructing of these wave
functions is closely related to the equations obtained in the
framework of the dispersion approach based on the analytic
properties of the scattering amplitudes ~\cite{KiT75,AnK92} (see
also ~\cite{KrT02} and especially the detailed version
~\cite{KrT01h}). In fact, this approach is a kind of dispersion
technique using integrals over composite--system masses.

Let us emphasize that by construction the dispersion wave functions
~\cite{MuT81} differ principially from deuteron wave functions used in the
conventional nuclear model (see, e.g., the review ~\cite{GiG02}). So,
for applications it is convenient to have MT--wave functions
~\cite{MuT81} in analytical form.

Therefore, we present here a  simple  perametrization of the
deuteron function as a superposition of Yukawa-type terms (that
was introduced in Ref.~\cite{LaL81} for Paris potential; see also
the fit in Ref.~\cite{Mac01} for CD-Bonn wave function).

So, we consider the deuteron wave functions $\varphi_l(r)$ in the
states with orbital momentum $\;l=0\;,\varphi_0(r)=u(r)$ and
$\;l=2\;,\varphi_2(r)=w(r)$
The ansatz for the analytic versions of the $r$-space wave
functions, denoted by $u_a(r)$ and $w_a(r)$, is
$$
u_a(r) = \sum\limits_{j=1}^{n_u}
{C_j}{\exp\left(-m_j\,r\right)}\;,
$$
\begin{equation}
w_a(r) = \sum\limits_{j=1}^{n_w} {D_j}{\exp\left(-m_j\,r\right)}
\left[1 + \frac{3}{m_j\,r} + \frac{3}{(m_j\,r)^2}\right]\;,
\label{uwr}
\end{equation}
$$
m_j = \alpha + m_0\,(j-1)\;,
$$
where the coefficients $C_j,\,D_j$, the maximal value of the
index $j$ and $m_0$ are defined by the condition of the best
fitting. $\alpha = \sqrt{M\,\varepsilon_d}$, $M$ is nucleon mass,
$\varepsilon_d$ is the binding energy of deuteron.

%%%%%%%%%%%%%%%%%%%%%%%%%%%%%%%%%%%%%%%%%%%%%
% Figure 1 wave function u(r)
%%%%%%%%%%%%%%%%%%%%%%%%%%%%%%%%%%%%%%%%%%%%%
\begin{figure}
 \includegraphics{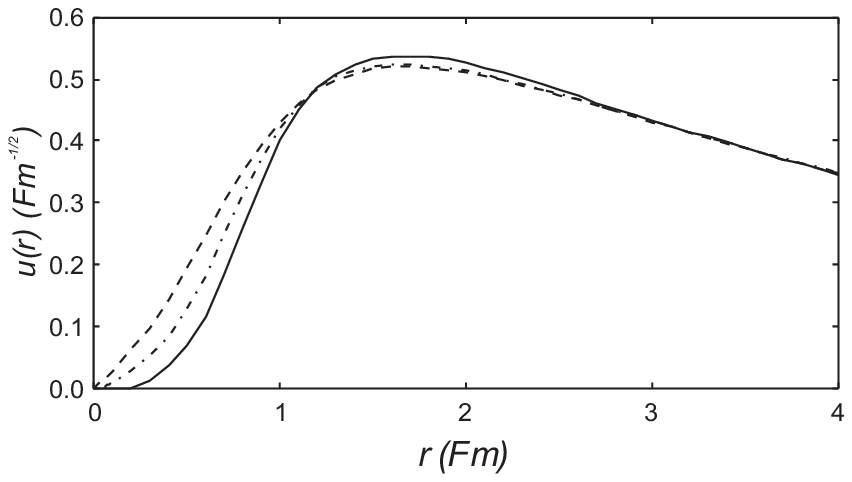}
 \caption{\label{figu} S-wave deuteron wave function. Solid line
- MT \protect\cite{MuT81,Tro94}, dushed --\protect\cite{Mac01},
dot-dushed -- \protect\cite{LaL81}.}
\end{figure}
%%%%%%%%%%%%%%%%%%%%%%%%%%%%%%%%%%%%%%%%%%%%%
% Figure 2 wave function w(r)
%%%%%%%%%%%%%%%%%%%%%%%%%%%%%%%%%%%%%%%%%%%%%
\begin{figure}
 \includegraphics{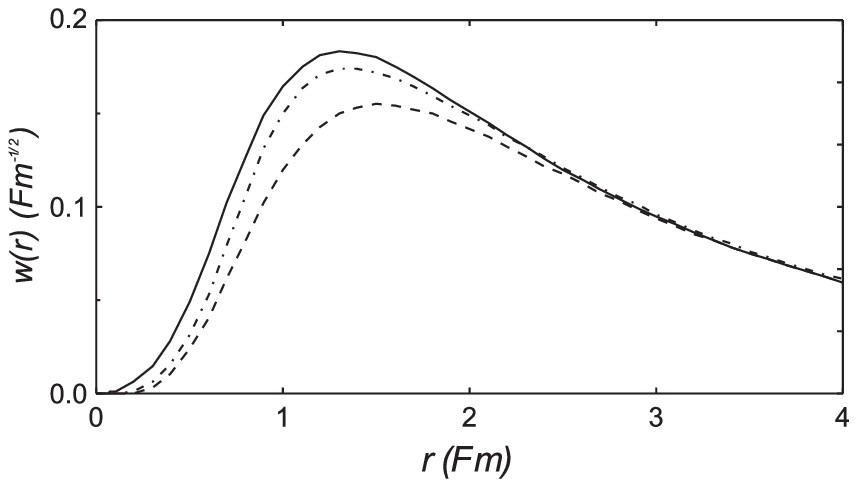}
 \caption{\label{figw} D-wave deuteron wave function. Legend is
the same as in Fig.\protect\ref{figu}.}
\end{figure}
%%%%%%%%%%%%%%%%%%%%%%%%%%%%%%%%%%%%%%%%%%%%%

These wave functions are normalized according to
\begin{equation}
\int_0^\infty\,dr\left[\left(u(r)\right)^2 +
\left(w(r)\right)^2\right] = 1\;. \label{normr}
\end{equation}
%%%%%%%%%%%%%%%%%%%%%%%%%%%%%%%%%%%%%%%%%%%%%%%%%%%%%%%%%%
\begin{table}
\caption{\label{tab:1}Coefficients for the parametrized deuteron
wave function calculated within a dispersion approach.  The last
$C_j$ and last three $D_j$ are to be computed from eq.
(\ref{cond})($n_u$=16, $n_w$=12).}
\begin{ruledtabular}
\begin{tabular}{ll}
%  Heading 1 & Heading 2\\
%  Cell 1 & Cell 2\\
$C_j$(fm$^{-1/2}$)&$D_j$(fm$^{-1/2}$)\\
          ~0.87872995+00        &    ~0.22245143-01\\
          -0.50381047+00        &    -0.41548258+00\\
          ~0.28787196+02        &    -0.18618515+01\\
          -0.82301294+03        &    ~0.21987598+02\\
          ~0.12062383+05        &    -0.16885413+03\\
          -0.10574260+06        &    ~0.76001430+03\\
          ~0.59534957+06        &    -0.22287203+04\\
          -0.22627706+07        &    ~0.43330023+04\\
          ~0.59953379+07        &    -0.54072021+04\\
          -0.11282284+08        & eq. (\ref{cond})         \\
          ~0.15181681+08        & eq. (\ref{cond})         \\
          -0.14519973+08        & eq. (\ref{cond})         \\
          ~0.96491938+07        &                          \\
          -0.42403857+07        &                          \\
          ~0.11092702+07        &                          \\
           eq. (\ref{cond})     &                          \\
\end{tabular}
\end{ruledtabular}
\end{table}
%%%%%%%%%%%%%%%%%%%%%%%%%%%%%%%%%%%%%%%%%%%%%%%%%%%%%%%%%%
The conventional boundary conditions at zero:
\begin{equation}
u(r)\;\sim\;r\;,\quad w(r)\;\sim r^3\;, \label{cond0}
\end{equation}
lead to one condition for $C_j$ and three constraints for $D_j$,
namely:
\begin{equation}
\sum\limits_{j=1}^{n_u} {C_j} = 0\;,\quad \sum\limits_{j=1}^{n_w}
{D_j} = \sum\limits_{j=1}^{n_w} {D_j}{m_j^2} =
\sum\limits_{j=1}^{n_w} \frac{D_j}{m_j^2} = 0\;. \label{cond}
\end{equation}

Using the form (\ref{uwr}) it is easy to describe the standard behaviour
of the deuteron wave functions at
$r\to\infty$. The asymptotics of $S$-state is
\begin{equation}
u(r)\;\sim\;A_S\,\hbox{e}^{-\alpha\,r}\;,
\label{As}
\end{equation}
and the asymptotics of $D$ state is
\begin{equation}
w(r)\;\sim\;\eta\,A_S\,\left(1 + \frac{3}{\alpha\,r} +
\frac{3}{\left(\alpha\,r\right)^2}\right) \hbox{e}^{-\alpha\,r}\;.
\label{etaAs}
\end{equation}
Here $A_S$ and $A_D = \eta\,A_S$ are the asymptotic $S$-state and
$D$-state normalizations and $\eta$ is the asymptotic ``$D/S$
state ratio''.  In our calculation of MT-wave-function we use
$\alpha$=0.231625 fm$^{-1}$.

The Fourier-transforms of wave functions $\psi_l(k)\;,\;l=0,2$ in the
momentum representation in $r$-space are:
\begin{equation}
\frac{\varphi_l(r)}{r} = \sqrt{\frac{2}{\pi}}
\int_0^\infty\,k^2dk\,j_l(kr)\psi_l(k)\;, \label{Ftransform}
\end{equation}
where $j_l(kr)$ is the spherical Bessel function.

The normalization condition for these wave functions is given by
\begin{equation}
\int_0^\infty\,k^2dk\left[\left(\psi_0(k)\right)^2 +
\left(\psi_2(k)\right)^2\right] = 1\;. \label{normk}
\end{equation}

The fits for momentum space wave functions, following from
(\ref{uwr}) and (\ref{Ftransform}) are
$$
\psi^a_0(k)=\sqrt{\frac{2}{\pi}}\sum\limits_j
\frac{C_j}{(k^2+m_j^2)}\;,
$$
\begin{equation}
\psi^a_2(k)=\sqrt{\frac{2}{\pi}}\sum\limits_j
\frac{D_j}{(k^2+m_j^2)}\;. \label{psi02k}
\end{equation}

The calculated coefficients in the fits (\ref{uwr}),
(\ref{psi02k}) are listed in the table 1.

The asymptotics at $r\;\to\;\infty$ gives for the obtained fits
of MT-wave functions the asymptotic ``$D/S$ state ratio''
\begin{equation}
\eta = \frac{D_1}{C_1} = 0.02531511\;.
\label{eta}
\end{equation}

The accuracy of this parametrization is illustrated by the magnitudes
of the integrals:
\begin{equation}
\left\{\int_0^\infty\,dr\left[u(r) -
u_a(r)\right]^2\right\}^{1/2} = 4.1\cdot 10^{-3}\;, \label{accur0}
\end{equation}
\begin{equation}
\left\{\int_0^\infty\,dr\left[w(r) -
w_a(r)\right]^2\right\}^{1/2} = 2.2\cdot 10^{-3}\;. \label{accur2}
\end{equation}

So, we present a convenient analytical parametrization of the
deuteron wave function calculated within dispersion approach as a
discrete superposition of Yukawa-type functions. This function is
plotted on the Fig.1 and Fig.2. The wave functions \cite{LaL81}
and \cite{Mac01} are given for comparison.


\begin{thebibliography}{99}

\bibitem{KrT06} A.F.~Krutov and V.E.~Troitsky, nucl-th/0607026.


\bibitem{MuT81} V.M.~Muzafarov and V.E.~Troitsky,
                Yad. Fiz. \textbf{33}, 1461 (1981)
                [Soviet J. Nucl.Phys. \textbf{33}, 783 (1981)].


\bibitem{Tro94} V.E.~Troitsky, in \textit{ Proceedings of Quantum
                Inversion Theory and Applications}, Germany, 1993, edited
                by H.V. von Geramb, Lecture Notes in Physics \textbf{467}
                (Springer, Berlin, 1994), p. 50.

\bibitem{KrT03} A.F.~Krutov and V.E.~Troitsky,
                Eur. Phys. J. A \textbf{ 16}, 285 (2003).

\bibitem{DrS04}D.~Drakoulakos et al.,
               {\it Proposal to Perform a High-Statistic Neutrino
               Scattering Experiment Using a Fine-graned
               Detector in the NuMI Beam}, hep-ex/0405002

\bibitem{KiT75} A.I.~Kirillov, V.E.~Troitsky, S.V.~Trubnikov, and
                Yu.M.~Shirokov,
                Fiz. Elem. Chastits At. Yad. \textbf{6}, 3 (1975)
                [Sov. J. Part. Nucl. \textbf{6}, 3 (1975)]

\bibitem{AnK92} V.V.~Anisovich, M.N.~Kobrinsky, D.I.~Melikhov, and
                A.V.Sarantsev,
                Nucl. Phys. A \textbf{544}, 747 (1992)

\bibitem{KrT02} A.F.~Krutov and V.E.~Troitsky,
                Phys. Rev. C \textbf{65}, 045501 (2002)

\bibitem{KrT01h} A.F.Krutov and V.E.Troitsky, hep-ph/0101327


\bibitem{GiG02} R.~Gilman and F.~Gross,
                J. Phys. G  \textbf{28}, R37 (2002).


\bibitem{LaL81} M.~Lacomb, B.~Loiseau, R.~Vinh Mau, J.~Cot\'e, P.~Pir\'es,
                and R.~de~Tourreil,
                Phys. Lett. B \textbf{101}, 139 (1981).

\bibitem{Mac01} R.~Machleidt,
                Phys. Rev. C \textbf{63}, 024001 (2001).



\end{thebibliography}
\end{document}